# A Stress-Induced Tetragonal Distortion in $K_2OsCl_6$ under Pressure

Neha Bura[1], Peijie Zhang[1], Ganesh Bera[1], Catalin Popescu[2], Caleb J. Bennett[3], Brendan J. Kennedy[3], and Daniel Errandonea[1]

[1]Departamento de Física AplicadaInstituto de Ciencia de Materiales, Matter at High Pressure (MALTA) Consolider Team, Universidad de Valencia, Edificio de Investigación, C/Dr Moliner 50, Burjassot, Valencia 46100, Spain

[2]CELLS-ALBA Synchrotron Light Facility, Cerdanyola, del Vallès, Barcelona 08290, Spain

[3]School of Chemistry, The University of Sydney, Sydney, New South Wales 2006, Australia

**Abstract**

The structural behaviour of cubic ($Fm\bar{3}m$) $K_2OsCl_6$ under compression was investigated to 35.4 GPa using angle-dispersive synchrotron powder X-ray diffraction. Upon increasing pressure, $K_2OsCl_6$ undergoes a sequence of structural transformations from the ambient-pressure cubic phase to a tetragonal ($P4/mnc$) phase above 9.5(3) GPa, followed by a transition to a monoclinic ($P2_1/n$) phase beginning at approximately 14.2 GPa. The tetragonal and monoclinic phases coexist over the pressure range 14.2-17.2 GPa, while the monoclinic phase is observed as the sole phase at 17.9 GPa and remains stable up to the maximum pressure investigated. Analysis of the diffraction profiles and unit-cell parameters indicates that the intermediate tetragonal structure is likely associated with stress-induced symmetry lowering rather than a distinct thermodynamic phase. This distortion, which is adequately described by the tetragonal $P4/mnc$ space group, is attributed to non-hydrostatic deviatoric stresses arising from the pressure-transmitting medium. Equation-of-state analysis reveals a pronounced reduction in compressibility across the structural transitions, with the bulk modulus increasing from 26(2) GPa for the cubic phase to 71(3) GPa for the monoclinic phase. The pressure-induced structural sequence is analogous to that reported for related compounds such as $K_2IrCl_6$, although the transitions in $K_2OsCl_6$ occur at lower pressures. All structural transformations are reversible upon decompression.

## Introduction

Halide perovskites encompass a broad range of compounds that display a rich array of structural, electronic, magnetic, and optoelectronic properties.[1-4] The discovery of the exceptional optoelectronic properties of lead halide perovskites, particularly the seminal demonstration by Miyasaka and co-workers in 2009 of their potential for photovoltaic applications, has stimulated extensive research of lead halide perovskites.[2,5] However, even the most promising hybrid organic–inorganic lead halide perovskites of the type $APbX_3$, where $A$ is a monovalent organic or inorganic cation and $X$ is a halide, are limited by the toxicity of Pb, their moisture sensitivity and poor stability. This has motivated extensive efforts to develop less toxic and more stable alternatives.[6, 7]

Cation-vacancy-ordered double perovskites (VODPs) of the type $A_2MX_6$ have emerged as a versatile platform for developing lead-free alternatives to conventional lead halide perovskites, as they offer numerous opportunities to tune their structural, electronic, optical, and magnetic properties.[8] Understanding their structural response to external stimuli is an important step in establishing general structure–property relationships. Studies of VODP span several decades and earlier work often focused on their thermally induced structural phase transitions and, when $M$ is a transition metal, their magnetic properties.[9-13] Contemporary studies of several VODPs containing heavier $5d$ transition metals, especially Ir and Os have demonstrated these halides can display unusual magnetic properties due to a delicate balance between competing structural instabilities and the coupling between the lattice degrees of freedom and the strongly spin–orbit-coupled $M^{4+}$ electronic state.[14]

The VODP $K_2OsCl_6$ and $K_2IrCl_6$ are members of the $K_2PtCl_6$ family of rock salt ordered double perovskite of the type $A_2MM'X_6$ where the $M'$ site is vacant. This structure, that is also described as an antifluorite type, features isolated $MX_6$ octahedra separated by alkali metal cations.[15, 16] The $MX_6$ octahedra are arranged on a face centre cubic lattice that leads to a geometrically frustrated magnetic lattice. The magnetic properties of the three $K_2MCl_6$ ($M$ = Re, Os, Ir) compounds exhibit remarkably different behaviour. $Re^{4+}$ has a $5d^3$electron configuration resulting in a S =3/2 ground state in an octahedral field.[17, 18] In $K_2ReCl_6$ Hund's coupling and exchange splitting are sufficiently strong, relative to the spin-orbit coupling (SOC), that the observed magnetic behaviour is predominantly spin-only and experimental studies have demonstrated that it shows Type-I antiferromagnetic ordering below about 11.9 K.[19] This is fundamentally different to the situation in $K_2OsCl_6$, and $K_2IrCl_6$ where SOC effects dominate. In the strong SOC limit, the $Os^{4+}$ $t_{2g}^4$ ion is expected to form a nonmagnetic $J_{\mathrm{eff}} = 0$ singlet state. Variable temperature magnetic susceptibility studies, however, show that $K_2OsCl_6$ is not a simple $J_{\mathrm{eff}} = 0$ Van Vleck paramagnet, rather it exhibits Kotani-like susceptibility behaviour with evidence for a weakly magnetic ground state.[20] In $K_2IrCl_6$ the strong SOC results in a Jeff = ½ ground state and $K_2IrCl_6$ shows Type-III AFM ordering at about 3 K.[21]

The Goldschmidt-type tolerance factor for the vacancy-ordered structure, that ignores the effect of the vacancies at the *M*' site, is given by the expression

$$t = \frac{r_A + r_X}{\sqrt{2}(r_M + r_X)}$$

where $r_i$ is the ionic radii of the appropriate ion. The structures of the three $5d$ halides $K_2MCl_6$ (*M* = Re, Os, and Ir) show remarkably different thermal properties despite the size of the $M^{4+}$ cation, and hence Goldschmidt-type tolerance factor being approximately equal. According to Shannon $r_M$ = 0.63 Å, for $Re^{4+}$ and $Os^{4+}$ and 0.625 Å for $Ir^{4+}$.[22] All three compounds have $t \sim 1.00$ suggestive of excellent geometrical compatibility with the parent cubic perovskite topology.

Both $K_2ReCl_6$ and $K_2OsCl_6$ show symmetry lowering transitions upon cooling associated with tilting of the essentially rigid $MCl_6$ octahedra whereas $K_2IrCl_6$ remains cubic to low temperatures. The observed sequence of structural transitions, $Fm\bar{3}m \rightarrow P4/mnc \rightarrow I2/m \rightarrow P2_1/n$ for *M* = Re[23] simplifies to $Fm\bar{3}m \rightarrow I4/m$ for M = Os[16] and appears to be more strongly influenced by the $M^{4+}$ electronic configuration; $Re^{4+}$ is $5d^3$, $Os^{4+}$ is $5d^4$, and $Ir^{4+}$ is $5d^5$, than their Goldschmidt tolerance factors. It should be noted that a recent neutron diffraction study demonstrated that the earlier report that the low temperature structure of $K_2OsCl_6$ was tetragonal in *P*4/*mnc* was incorrect.[16] The *P*4/*mnc* and *I*4/*m* structures differ in the nature of the cooperative tilts of the octahedra. In *P*4/*mnc* the $MX_6$ octahedra have in-phase tilting about the c-axis ($a^0a^0c^+$ in Glazer's notation) whereas in *I*4/*m* they rotate in the opposite sense ($a^0a^0c^-$). Total-scattering studies of $K_2IrCl_6$ have shown that although the average crystallographic structure remains cubic down to 11 K, it has local monoclinic $P2_1/n$-like distortions, involving both in-phase and out-of-phase octahedral tilts.[24]

We recently demonstrated that the local scale structure in $K_2IrCl_6$ can be stabilised as the long-range or average structure by the application of applied pressure.[24] High-pressure synchrotron measurements revealed that $K_2IrCl_6$ undergoes two structural transitions: cubic $Fm\overline{3}m$ to tetragonal $P4/mnc$ at 12.0 GPa, and tetragonal to monoclinic $P2_1/n$ at 15.1 GPa. In the present work, we extend our high-pressure study of $K_2IrCl_6$ to $K_2OsCl_6$ using synchrotron X-ray diffraction to 35.4 GPa. We are interested to explore if there is a correlation between the $M^{4+}$ electronic configuration and the observed pressure induced structures or if the pressure induced transitions are more strongly influenced by geometric factors. $K_2OsCl_6$ therefore offers a useful platform for exploring pressure as a means of tuning the interplay between structure, spin–orbit coupling, and magnetism in molecularly isolated $5d^4$ transition-metal systems. We found that $K_2OsCl_6$ exhibits the same overall cubic-to-tetragonal-to-monoclinic structural sequence observed in $K_2IrCl_6$, but at lower pressures.

**Experimental Procedure**

A sample of $K_2OsCl_6$ was purchased from Surepure Chemicals and used as received. The high-pressure XRD studies on $K_2OsCl_6$ were carried out at the MSPD Beamline of ALBA facility. Angle-dispersive synchrotron X-ray diffraction (AD-XRD) data were collected for pressures of up to 35.4 GPa. The wavelength of the monochromatic radiation was determined to be 0.4246 Å. XRD data were collected using a Rayonix SX165 CCD detector. High-pressure measurements were carried out using a membrane-driven diamond-anvil cell (DAC) with 300 μm culets. A Rhenium gasket with an initial thickness of 200 μm was pre-indented to a thickness of 40 μm, and a 200 μm hole was drilled at the center to act as the sample chamber. The sample was loaded into the chamber along with Cu as an internal pressure calibrant. The pressure was determined from the Cu (111) reflection using its known EOS[25] with an accuracy of 0.5%. Fluorinert FC-70, having a hydrostatic limit of around 1-2 GPa,[26] was used as the pressure-transmitting medium (PTM). During data acquisition the DAC was rotated +/- 3º to improve the quality of the diffraction data by reducing preferred orientation and grain-growth issues caused by high pressure. The sample-detector distance was calibrated using a $LaB_6$ standard. Conventional 1D diffraction patterns were obtained by integrating the detected intensity as a function of 2θ using Dioptas. Prior to integration saturated pixels were masked.

The structure analysis was undertaken using the Rietveld method as implemented in the GSAS software package.[27] The diffraction profiles were modelled using a pseudo-Voigt peak-shape function, and the background was described by a $12^{th}$-order Chebyshev polynomial. During the fitting, the scale factor, lattice parameters, isotropic displacement parameters, and peak-profile parameters were refined. Atomic site occupancy factors were fixed at the nominal composition. In the final refinement cycles, all unconstrained parameters were allowed to vary simultaneously, and the refined crystal structures were visualized using VESTA.[28]

## Results

Figure 1 presents pressure-dependent XRD profiles for $K_2OsCl_6$ measured at selected pressures of up to 35.4 GPa. As the pressure increases the diffraction peaks move towards higher values of 2θ because of compression of the lattice. As evident from this figure, the sample transitions through various structural phases as the applied pressure is increased. All the peaks in the lowest pressure profile could be indexed to a F-centered cubic structure (space group $Fm\bar{3}m$). This cubic structure was retained at pressures of up to 9.2 GPa. The unit-cell parameter changes from 9.6176(9) Å at 0.9 GPa to 8.9974(9) Å at 9.2 GPa, a 6.5% contraction. Upon further compression, at approximately 9.8 GPa the structure transforms to a high-pressure phase, which we named HP1. This phase was identified as being tetragonal and described in space group $P4/mnc$ with lattice parameters $a$ = 6.3023(6) Å and $c$ = 9.0613(9) Å at 9.8 GPa. We define the transition pressure as the intermediate pressure between the last observation of the cubic structure and the first observation of the tetragonal structure; i.e. 9.5(3) GPa. Apparently, there is no discontinuity in the volume at the transition. The absence of a measurable volume discontinuity at the cubic-to-tetragonal transition argues against a first-order transformation and is consistent with either a continuous structural instability or a stress-induced tetragonal distortion.[29] This tetragonal phase was observed in the experiments to 13.5 GPa. Further increasing the pressure leads to a more complex diffraction profile at 14.2 GPa that resulted from the progressive decrease in intensity of reflections from HP1 and the emergence of new reflections from a second high-pressure phase, which we named HP2. These two phases co-existed at pressures of up to 17.2 GPa. There was no evidence for any peaks associated with the HP1 phase at or above 17.9 GPa and the resulting profiles were consistent with a monoclinic structure described in space group $P2_1/n$. The HP2 phase was found to persist to the maximum studied pressure of 35.4 GPa. The lattice parameters at highest studied pressure are $a$ = 6.7817(9) Å, $b$ = 6.0473(9) Å, $c$ = 7.2055(9) Å, and β = 93.92(2)°. At the tetragonal-to-monoclinic transition there is an apparent volume discontinuity of 2.7% indicating the transition is first order. The crystal structures were refined by the Rietveld method and representative examples of such refinements are shown in the Figure 2. The CIF files of the tetragonal and monoclinic structures have been deposited with the Cambridge Crystallographic Data Centre (CCDC) and are available under deposit number 2587907 and 2587908, respectively.

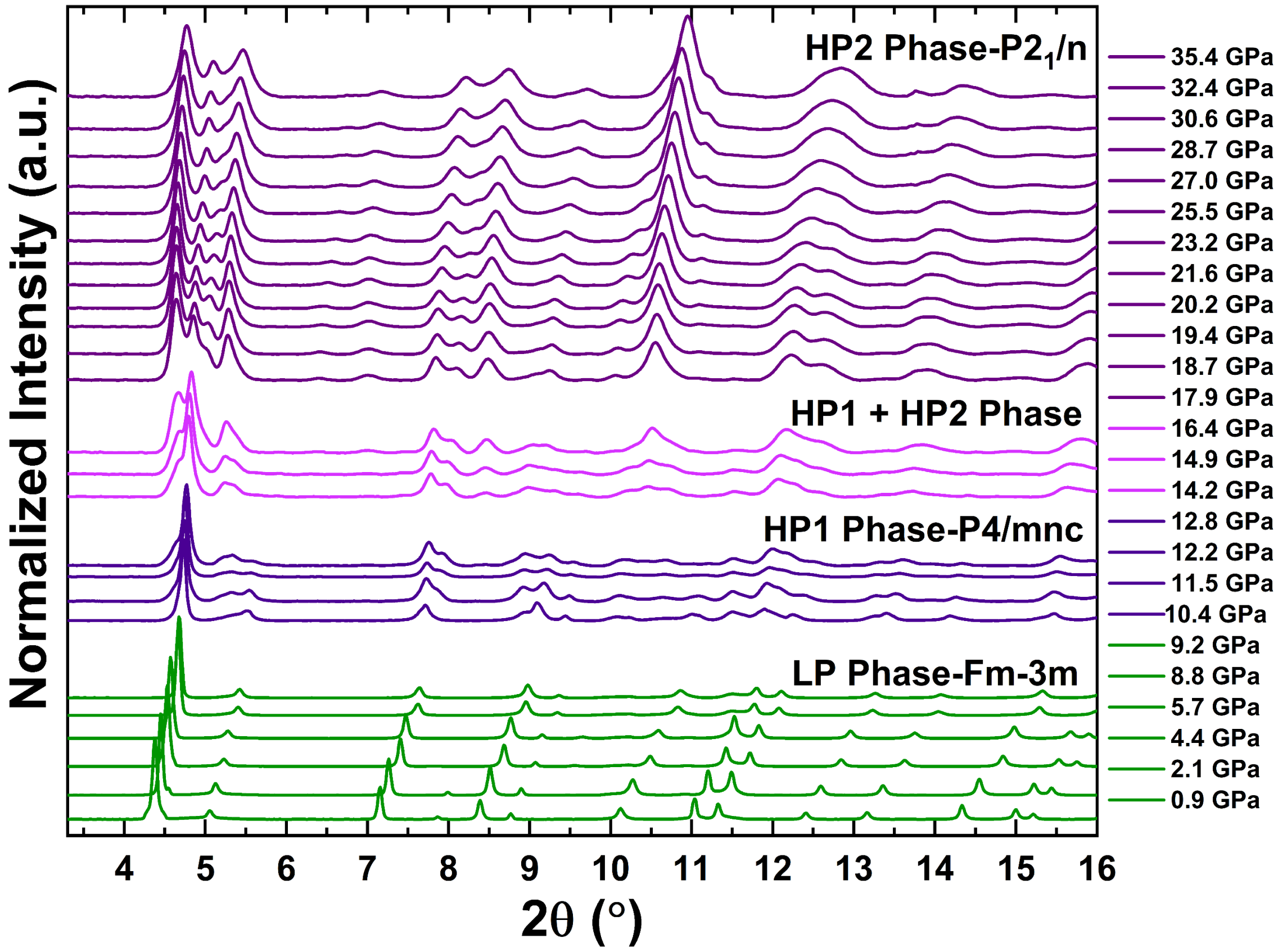

Figure 1. Pressure-dependent evolution of the powder X-ray diffraction patterns of $K_2OsCl_6$ ($\lambda$ = 0.4246 Å). We show selected pressure points with the intensity plotted on a common scale for clarity. The diffraction patterns show the structural transformation from the cubic phase at low pressure to the intermediate tetragonal state (HP1). In the figure, XRD patterns from 0.9 to 9.2 GPa are identified with the low-pressure phase and thos from 10.4 to 12.8 GPa are identified with the tetragonal HP1 phase. There is a second transition to the monoclinic HP2 phase. XRD patterns from 14.2 to 16.4 GPa show a coexistence of phases HP1 and HP2. From 17.9 to 35.4 GPa XRD patterns can be assigned to the monoclinic HP2 phase.

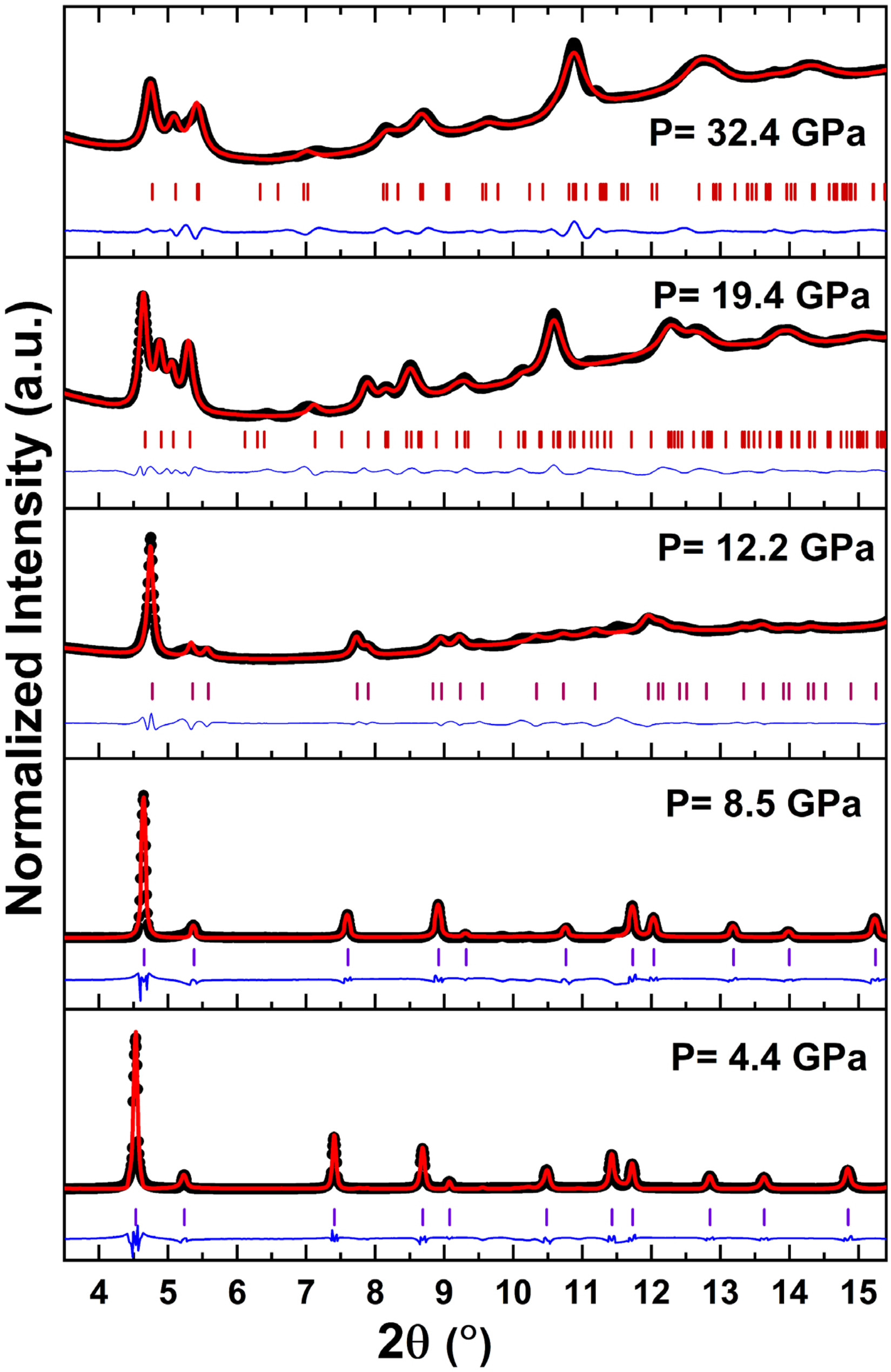

Figure 2. Representative Rietveld fits to AD-XRD data for $K_2OsCl_6$ collected using $\lambda$ = 0.4246 Å. The data measured at 4.4 and 8.5 GPa were fit using the cubic, $Fm\bar{3}m$, model, at 12.2 GPa using a tetragonal $P4/mnc$ model, and at 19.4 and 32.4 GPa a $P2_1/n$ model. The black symbols, and red and blue lines, represent the experimental data, calculated and difference profiles, respectively. The positions of the space group allowed Bragg reflections are indicated by the purple, pink and red bars, immediately below the XRD profiles. The Rp values for the fits at 4.4, 8.5, 12.2, 19.4, and 32.4 GPa are 1.98%, 3.91%, 15.8%, 17.3%, and 22.57%, respectively.

The diffraction pattern of the sample measured upon decompression, shown in Figure 3, resembles that measured in the DAC prior to compression. There are only four differences. Upon decompression, the diffraction peaks are at higher angles compared with those measured at the initial low-pressure state. This shift is attributed primarily to the fact that the sample was not fully decompressed to ambient pressure; the minimum pressure reached upon decompression in the experiment was approximately 2 GPa. The residual pressure is associated with friction between the piston and cylinder of the diamond-anvil cell, which prevents complete pressure release.[30] Therefore, the higher-angle position of the diffraction peaks after decompression is consistent with the remaining compressive pressure on the sample and should not be interpreted as evidence for a residual structural contraction. The diffraction peaks also become broader after decompression, which may reflect residual deviatoric stress and micro-strain accumulated during the pressure cycle.[31, 32] In addition, changes in the relative intensities of the reflections are observed, likely resulting from pressure-induced preferred orientation of the crystallites.[33] Finally, there are few extra peaks in the profile measured after decompression which could be assigned either to the Re gasket or the Cu pressure standard. The recovery of the cubic phase demonstrates that both the cubic to tetragonal and tetragonal to monoclinic phase transitions are reversible.

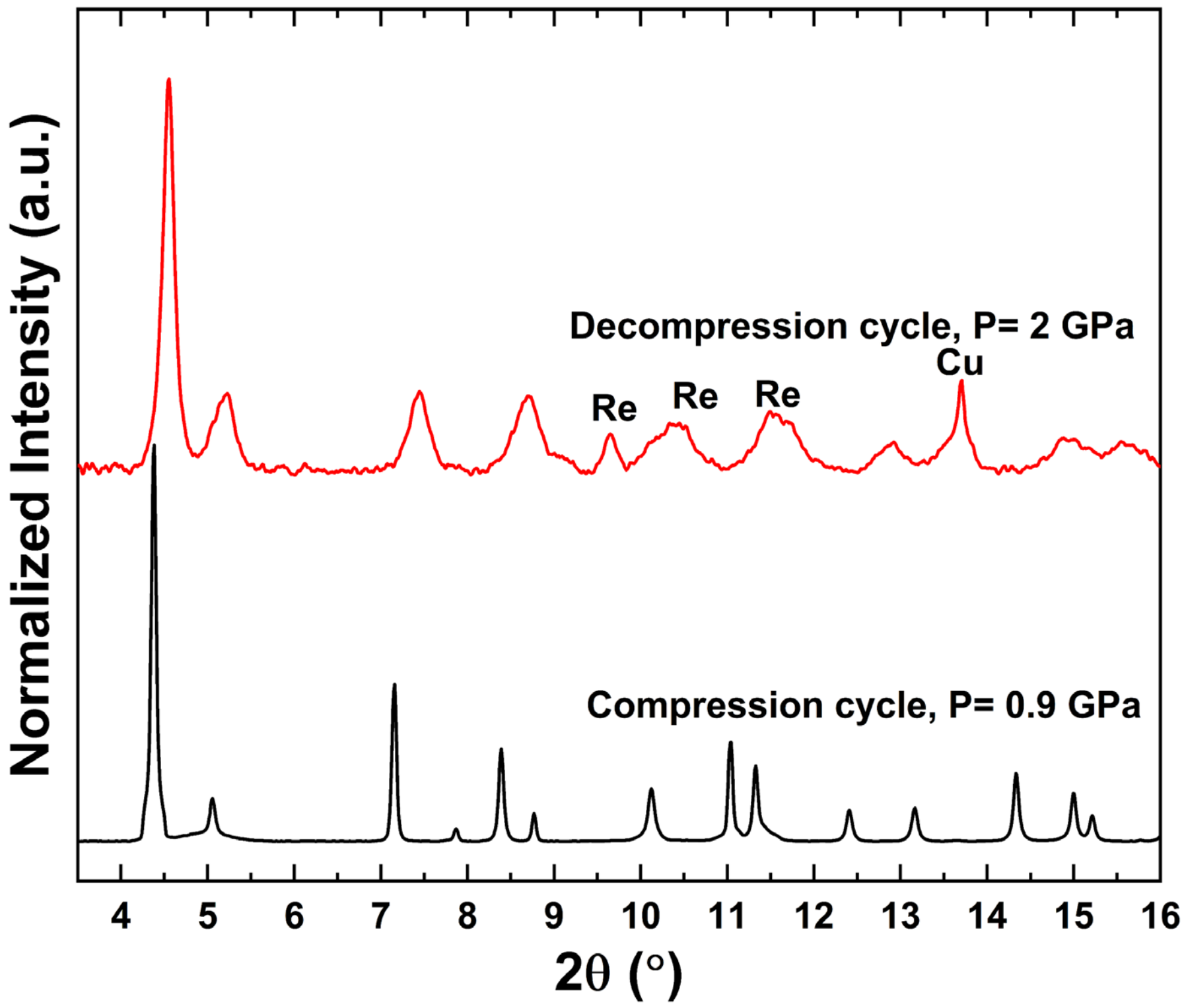

Figure 3. The comparison of the AD-XRD profiles for $K_2OsCl_6$ measured at the lowest pressure of compression and after decompression from 35.4 GPa ($\lambda$ = 0.4246 Å).

The sequence of phase transitions observed in $K_2OsCl_6$, $Fm\bar{3}m \leftrightarrow P4/mnc \leftrightarrow P2_1/n$, is consistent with the group theoretical analysis of Howard *et al.* and can be understood in terms of the successive introduction of cooperative tilting of the $OsCl_6$ octahedra. Using the Glazer notation this sequence is $a^0a^0a^0 \leftrightarrow a^0a^0c^+ \leftrightarrow a^-a^-c^+$ and the relationship between the lattice parameters of the low-pressure cubic phase and the high-pressure tetragonal phase can be expressed as $a_t \sim a_c/2$ and $c_t \sim a_c$, where the subscripts *t* and *c* refer to the tetragonal and cubic cells respectively. The pressure dependences of the appropriately scaled lattice parameters are shown in Figure 4. The cubic lattice parameter decreases by approximately 6.5% between 0.9 and 9.2 GPa, which is substantially larger than the 0.88% contraction observed upon cooling from 300 to 90 K. In the tetragonal phase, compression is strongly anisotropic, with the *a*-axis decreasing by 2.4%, while the *c* axis shows a slight expansion of approximately 0.1%. Such anisotropic lattice responses are characteristic of structural distortions associated with cooperative octahedral tilting and have been widely reported in perovskite-related systems.[34, 35] Similar anisotropic compression is observed in the monoclinic phase, where the *a*, *b*, and *c*-axes contract by approximately 1.9, 2.3, and 5.4%, respectively. In the monoclinic cell $a_m \sim b_m \sim a_c/\sqrt{2}$ and $c_m \sim a_c$ This implies that in the tetragonal phase, the reduction in the unit-cell volume is dominated by compression along the *a*-axis, whereas in the monoclinic phase it is predominantly governed by the substantially larger contraction along the *c*-axis. These contrasting anisotropic responses highlight the strong coupling between lattice compression and octahedral tilting across the pressure-induced structural transitions. The monoclinic angle, β, increases gradually with pressure, from approximately 92.5° at 17.9 GPa to 93.9° at 34.9 GPa. This progressive increase indicates a slight pressure-induced distortion of the crystal lattice, with the monoclinic angle becoming increasingly deviated from 90°. The change in β of 1.4° suggests that the monoclinic structure remains stable over this pressure range, while undergoing a gradual anisotropic structural deformation.

The co-existence of the tetragonal *P*4/*mnc* and monoclinic $P2_1/n$ phases between approximately 14.2 and 17.2 GPa is, at first sight, somewhat surprising as this is indicative of a first-order phase transition. Howard *et al.* showed that since this transition involves the introduction of a single out-of-phase tilt about (110) described by the irrep $R_4^+$ it is allowed to be continuous.[36] It is thought that this unexpected observation is a consequence of non-hydrostatic stresses resulting from the use of FC-70 as the PTM. The discontinuous volume change at the transition provides additional evidence for the first-order nature of the tetragonal to monoclinic transition.

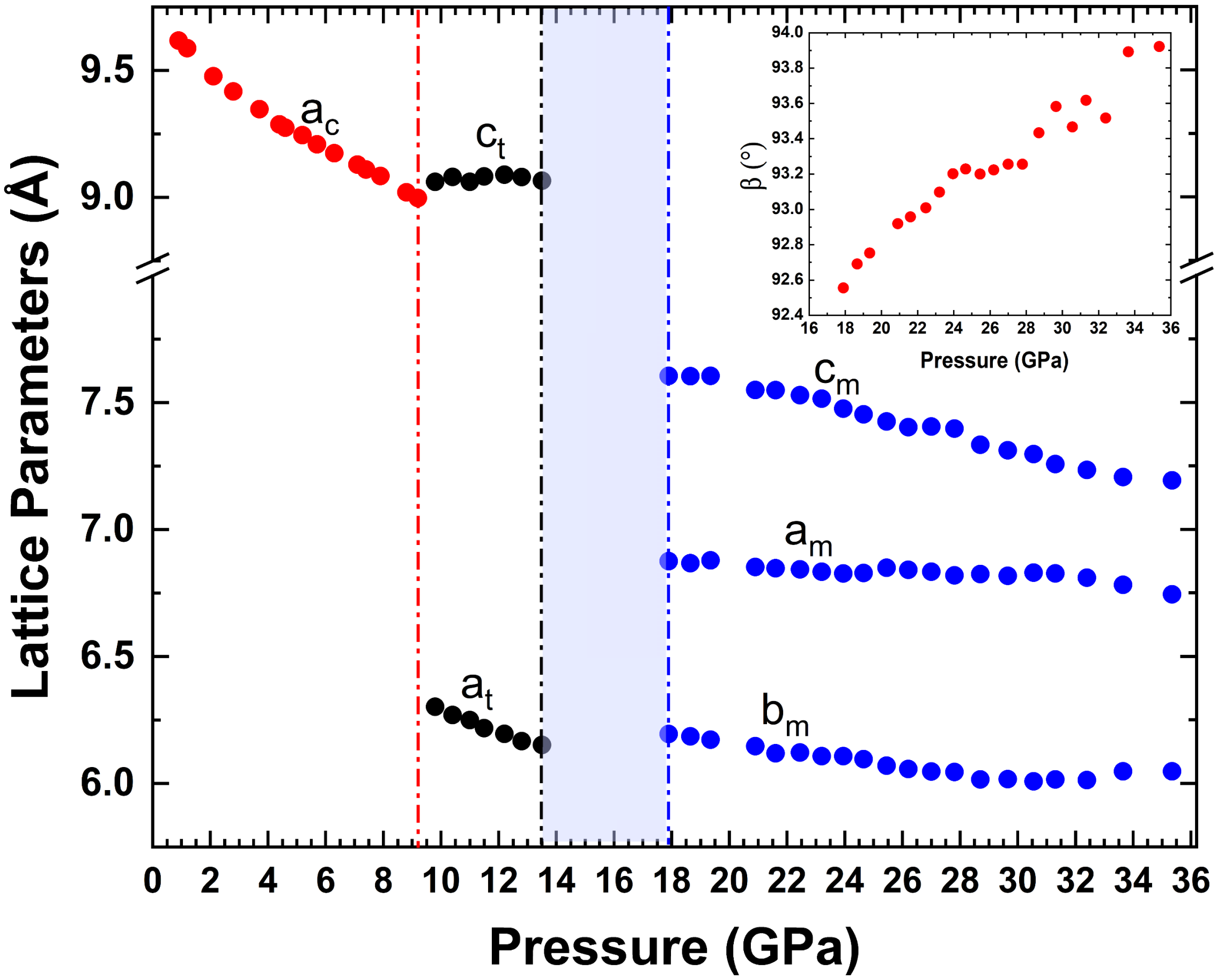

Figure 4. Pressure dependence of the appropriately scaled lattice parameters of $K_2OsCl_6$ obtained from HP AD-XRD measurements. The inset shows the variation of the monoclinic angle β with pressure. The increase in β with pressure indicates an increase in monoclinic distortion. The co-existence of the tetragonal and monoclinic phases between 14.2 and 17.2 GPa resulted in larger than optimal standard deviations for the refined lattice parameters, and for clarity such data are not included in this figure.

The Pressure-Volume (*P-V*) data were fitted with the third-order Birch-Murnaghan Equation of State (BM-EOS).[37] For the low-pressure cubic phase, the volume at zero pressure $V_0$, the bulk modulus at zero pressure $B_0$, and its pressure derivative $B_0'$ were refined. The results of the fitting and EOS parameters are shown in Figure 5 and in Table 1. For both HP phases, $B_0'$ was fixed at 3.9, as obtained for the low-pressure phase, and only $V_0$ and $B_0$ were fitted. Since $B_0$ and $B_0'$ are correlated,[38] it was decided to use the same value of $B_0'$ for the three phases to enable a meaningful, direct, comparison of the bulk modulus for each phase. Data points from the phase coexistence region were excluded from the equation-of-state (EOS) fit because the diffraction peaks strongly overlap, resulting in large uncertainties in the refined lattice parameters and significant pressure gradients within the transition region. The goodness of the fit of different EOSs to the experimental data for the three phases validates our approximation. For the EOS of the low-pressure phase, we got $\chi^2$ = 1.26, for HP1 $\chi^2$ = 1.47, and for HP2 $\chi^2$ = 1.78.

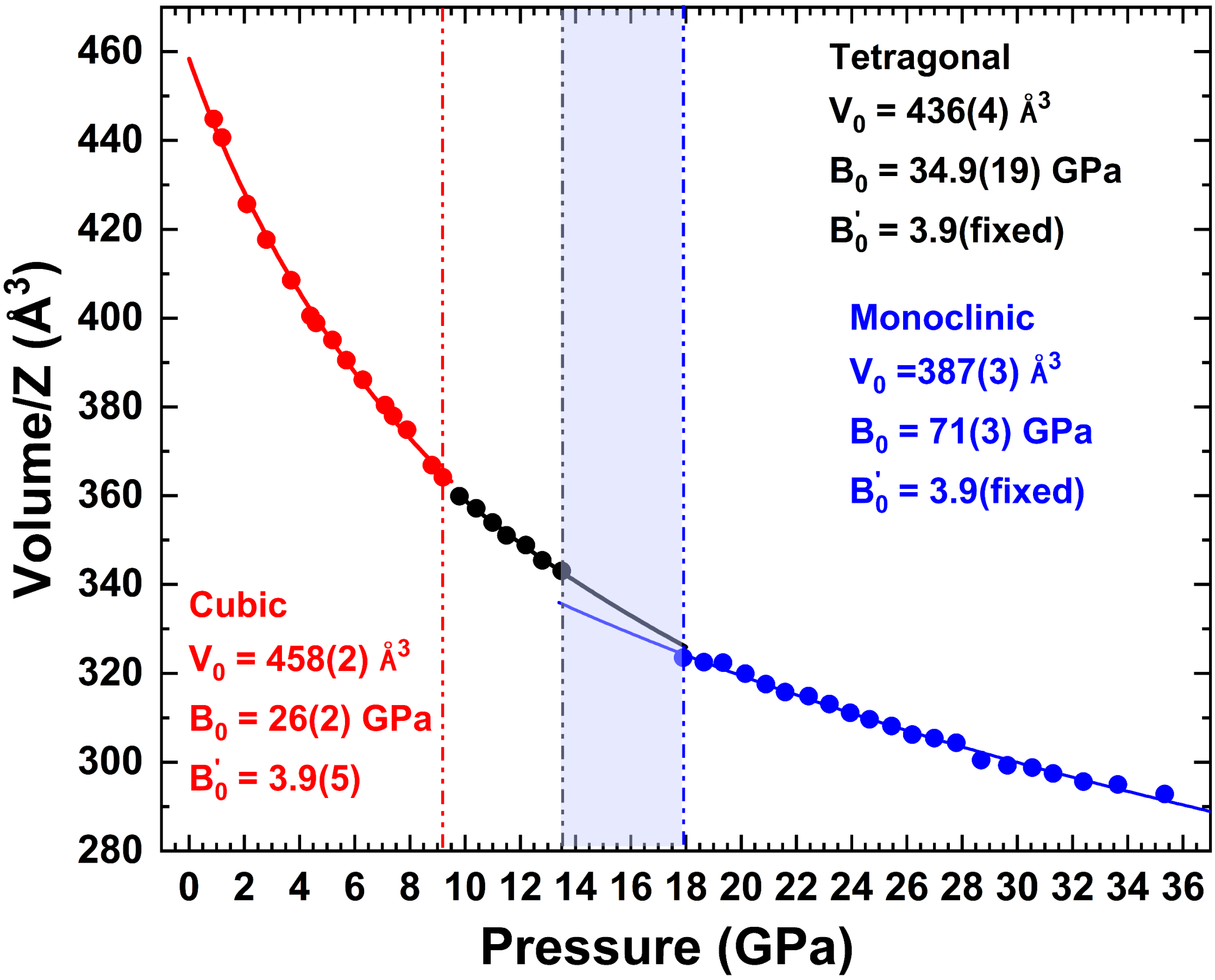

Figure 5. Variation of unit-cell volume for $K_2OsCl_6$ with pressure. Data points from the phase coexistence region were excluded from the equation-of-state (EOS) fit because the diffraction peaks strongly overlap, resulting in large uncertainties in the refined lattice parameters and significant pressure gradients within the transition region. Symbols are the experimental data and lines the EOS fits described in the text. The EOS of phases HP1 and HP2 have been extrapolated across the coexistence region.

Table 1 summarizes the $B_0$ values for $K_2OsCl_6$ calculated in the present study, together with the corresponding values previously reported for $K_2IrCl_6$. For $K_2OsCl_6$, the fitted $B_0$ increases from 26(2) GPa for the cubic structure to 34.9(19) GPa for the tetragonal structure and 71(3) GPa for the monoclinic structure. The bulk modulus of the low-pressure phase is in close agreement to that of the same phase in $K_2IrCl_6$,[24] KCl,[39] and other halide compounds.[40] The same agreement is observed when the tetragonal phases of $K_2IrCl_6$ and $K_2OsCl_6$ are compared (see Table 1). These observations can be attributed to the overall compressibility of their ionic lattices, which is governed by the collective response of interionic and inter-polyhedral interactions rather than solely by the strength of individual chemical bonds. The slightly lower bulk modulus of KCl (≈18 GPa) compared with $K_2IrCl_6$ (≈23 GPa) and $K_2OsCl_6$ (≈26 GPa) indicates that the latter compounds are somewhat less compressible. This behaviour can be attributed to the presence of relatively rigid $[MCl_6]^{2-}$ ($M$ = Ir, Os) octahedral units, together with their three-dimensional ionic lattice, which provide additional resistance to volume compression compared with the simpler KCl lattice. The higher bulk modulus of $K_2OsCl_6$ relative to $K_2IrCl_6$ may further reflect differences in metal–chlorine bonding, crystal geometry, and electronic structure, which collectively influence the stiffness and compressibility of the lattice.

Table 1. Comparison of bulk moduli for $K_2OsCl_6$ and $K_2IrCl_6$ across different structural phases. For the three phases of $K_2OsCl_6$ the pressure derivative of the bulk modulus is $B_0'$ is 3.9. For the three phases of $K_2IrCl_6$ the pressure derivative of the bulk modulus is $B_0'$ is 5.2. The results of $K_2IrCl_6$ are taken from the literature.[24]

| | Bulk Moduli $B_0$ (GPa) | | |
|---|---|---|---|
| Compound | Cubic | Tetragonal | Monoclinic |
| $K_2OsCl_6$ | 26(2) | 34.9(19) | 71(3) |
| $K_2IrCl_6$ | 23(2) | 33(2) | 121(5) |

The close agreement in the bulk moduli between the cubic and tetragonal phases suggests comparable lattice compressibility in these structures, whereas the substantially lower bulk modulus of the monoclinic $K_2OsCl_6$ phase indicates a markedly softer high-pressure structure than its Ir analogue. Since the ionic radii of sixfold-coordinated $Os^{4+}$ and $Ir^{4+}$ are very similar, this pronounced difference in the monoclinic phase is unlikely to originate from ionic size alone and may instead reflect differences in the metal-Cl bonding and structural response to compression. Evidently, the monoclinic structure has the lowest compressibility, followed by the tetragonal structure. The substantial increase in the bulk modulus following the second phase transition cannot be readily explained by simple chemical bond shortening alone. As the present diffraction data do not allow for isolating the contributions from pressure-induced changes in bond angles, polyhedral connectivity, or their coupling, the microscopic origin of this pronounced stiffening remains to be fully established through further structural analysis and theoretical calculations. At the cubic to tetragonal phase transition there is no detectable volume collapse within the accuracy of measurements, while the change in volume is 2.7% during tetragonal to monoclinic phase transition. This supports the idea that the second transition is first order as identified by the co-existence of the tetragonal HP1 and monoclinic HP2 phases between approximately 14.2 and 17.2 GPa. We note that the monoclinic phase of $K_2OsCl_6$ has a much smaller bulk modulus (71 GPa) than observed for the same phase in $K_2IrCl_6$ (121 GPa). This and the increase by four times of the bulk modulus at the tetragonal to monoclinic transition in $K_2IrCl_6$ suggest that the EOS of the monoclinic phase in this compound might be considerably affected by non-hydrostatic stresses.[41] Further studies using a selection of pressure media providing different degrees of non-hydrostaticity are required to fully understand the influence of deviatoric stresses in the HP structural sequence of $K_2IrCl_6$, $K_2OsCl_6$, and related chlorides.

**Discussion**

We will start the discussion by further elaborating on the identification of the crystal structure of the tetragonal phase. We hypothesize that the structure may adopt either the *I*4/*m* or *P*4/*mnc* space group. In our experiments, both models yield essentially the same lattice parameters and very similar diffraction patterns, with the main differences being limited to a few weak reflections because $h + k + l$ = odd is allowed in *P*4/*mnc* but not in *I*4/*m*. The difference between the refinements performed assuming both models can be seen in Figure 6. *P*4/*mnc* gives a better fit (Rp = 15.8%) than *I*4/*m* (Rp = 17.4%). More importantly, *P*4/*mnc* accounts for the additional weak reflections (indicated by arrows in Figure 6) that *I*4/*m* cannot explain. Based on these observations, we conclude that the tetragonal structure is best described using the *P*4/*mnc* model.

However, the crystallographic identification of *P*4/*mnc* does not by itself establish that this structure represents an independent thermodynamic phase. Our data indicate that deviatoric stress may stabilize, or enhance, the *P*4/*mnc* distortion, although the present measurements do not allow us to establish whether *P*4/*mnc* represents an independent thermodynamic phase. This possibility is considered further below in the context of the nonhydrostatic conditions associated with the FC-70 pressure-transmitting medium.

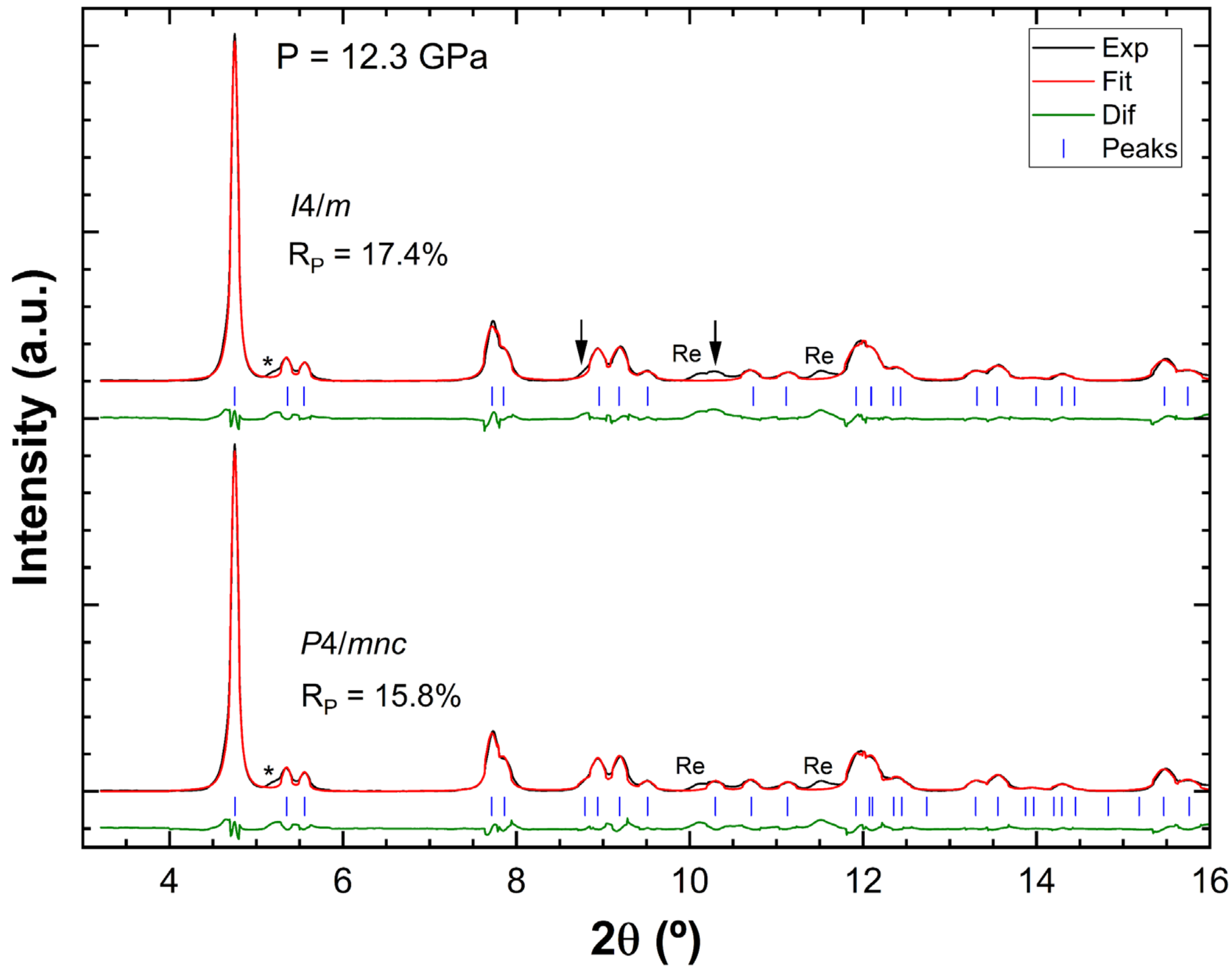

Figure 6. Rietveld refinement performed using the *I*4/*m* (top) and *P*4/*mnc* (bottom) models. Peaks from the rhenium gasket are identified as Re and a peak from solid FC-70 with *. The arrows identify two weak reflections that cannot be accounted by *I*4/*m*.

We now discuss the possible reasons for the observation of the *P*4/*mnc* structure under compression instead of the *I*4/*m* structure observed at low temperature. In FC-70, the hydrostatic pressure range extends up to approximately 2 GPa,[26] beyond which the pressure gradient increases rapidly, reaching ~0.5 at 10 GPa. The structural phase transitions in $K_2OsCl_6$ occur under pressure conditions that are no longer strictly quasi-hydrostatic, increasing the possibility that the observed pressure-induced changes are influenced by non-hydrostatic stresses, as demonstrated for $K_2OsCl_6$ under uniaxial pressure[42] and for several oxide systems.[43, 44] As shown in Figure 2, in the present case, the diffraction data provide evidence for the presence of deviatoric stresses. In particular, the strongest low-angle reflection, which is expected to remain a singlet in the tetragonal phase, exhibits pronounced broadening. Since no symmetry-related splitting of this reflection is anticipated, its broadening is more plausibly attributed to non-hydrostatic strain rather than to a crystallographic symmetry change. Moreover, the pressure dependence of diffraction broadening provides independent evidence that deviatoric stress becomes significant in the same pressure interval. Given the close structural relationship between the *I*4/*m* and *P*4/*mnc*, space groups, it is conceivable that deviatoric stresses distort the *I*4/*m* structure in a manner that can be adequately described by *P*4/*mnc*, without necessarily indicating a genuine thermodynamic phase transition. This interpretation is further supported by the refined unit-cell parameters, which remain essentially identical for the two structural models. We therefore suggest that the observed change is more appropriately described as a subtle stress-induced symmetry lowering of the (*I*4/*m*) structure rather than the formation of a distinct crystallographic phase. This interpretation is also consistent with the low-temperature neutron diffraction study of Saura-Múzquiz *et al.*,[16] who identified the tetragonal *I*4/*m* phase as an intermediate structure in the corresponding phase-transition sequence. This hypothesis could be confirmed by future experiments with Cryogenic DAC (which combines low-temperature and high-pressure) using a pressure-transmitting medium such as helium or neon that remain quasi-hydrostatic up to 20 GPa.

More generally, the tetragonal phase in this family of compounds occupies a relatively narrow region of the temperature-pressure phase space, and the transition from the tetragonal to the monoclinic structure is first order. From a group-theoretical perspective, a transformation between *I*4/*m* and *P*4/*mnc* is likewise expected to be first order,[36] and there is presently no obvious thermodynamic driving force for such a transition. The behaviour of $K_2OsCl_6$ appears to be somewhat unusual within this family as this contrasts with related compounds such as $K_2OsBr_6$, $K_2ReCl_6$, and $K_2SnCl_6$, for which the primitive tetragonal structure is observed upon cooling. [13,17,16]

The structural behaviour of $K_2OsCl_6$ suggests that the tetragonal phase may be particularly sensitive to non-hydrostatic conditions. The limited stability range of the tetragonal phase, together with the first-order character of its subsequent transformation to the monoclinic structure, indicates that relatively small deviations from hydrostatic compression could influence the structural response. In particular, deviatoric stresses may perturb the tetragonal lattice and promote local distortions, potentially giving rise to a structure that is better described by the *P*4/*mnc* model. First-principles calculations will be required to assess the relative stability of the competing tetragonal configurations and to determine whether such a stress-induced structural distortion is energetically favourable.

The sequence of structural phase transitions is similar to that observed for $K_2IrCl_6$ i.e.,

cubic→tetragonal→monoclinic

However, the transition pressures, especially that to the monoclinic phase, are much lower than in $K_2IrCl_6$. Given the small difference in ionic radii of 6-coordinate $Os^{4+}$, 0.63 Å, and $Ir^{4+}$, 0.625 Å, this difference in behaviour points to the importance of electronic, rather than geometric, factors. Specifically, this variation stems from the distinct 5*d* configurations under strong spin-orbit coupling. For $Ir^{4+}$ ($5d^5$), the half-filled ($J_{eff}$ = 1/2) doublet forms a symmetric ground state that stabilizes the cubic lattice against compression. In contrast, $Os^{4+}$ ($5d^4$) forms a non-magnetic ($J_{eff}$ = 0) singlet state. The lack of an open-shell ($J_{eff}$ = 1/2) state alters the potential energy landscape, enhancing the coupling between electronic states and the soft rotary phonon modes responsible for octahedral tilting. These differences suggest that the distinct 5d electronic configurations and associated spin–orbit-coupled electronic structures may contribute to the different pressure scales of the structural instabilities. Establishing the microscopic origin of this difference will require complementary first-principles calculations.

**Conclusion**

In conclusion, this high-pressure synchrotron X-ray diffraction study of $K_2OsCl_6$ reveals a reversible cubic ($Fm\bar{3}m$) to tetragonal ($P4/mnc$) to monoclinic ($P2_1/n$) structural sequence driven by applied pressure. The material transitions to the tetragonal phase at 9.5(3) GPa, followed by a robust transformation to the stable monoclinic structure at 14.2 GPa. Notably, our analysis indicates that the intermediate tetragonal state is strongly influenced by deviatoric stress introduced by the non-hydrostatic FC-70 pressure medium. The structural similarity between the *P*4/*mnc* and *I*4/*m* descriptions, combined with pronounced strain broadening, suggests that this intermediate state may stem from a stress-induced symmetry lowering of the I4/*m* structure rather than representing a distinct thermodynamic phase.

These structural transformations are associated with cooperative octahedral tilting and highly anisotropic lattice compression, with no detectable volume discontinuity at the cubic-to-tetragonal transition and an apparent volume discontinuity of approximately 2.7% at the tetragonal-to-monoclinic transition, accompanied by a significant stiffening of the lattice as the bulk modulus increases from 26(2) GPa to 71(3) GPa, indicating progressive stiffening of the lattice across the pressure-induced transitions.

While this sequence mirrors the behavior of $K_2IrCl_6$, the lower transition pressures in $K_2OsCl_6$ highlight the role of *d*-electron configuration and metal–Cl bonding over mere ionic size. Ultimately, future cryogenic diamond-anvil cell experiments utilizing highly quasi-hydrostatic media, such as helium or neon, are essential to definitively isolate the thermodynamic status of the tetragonal phase from stress-induced distortions.

### Author Contributions

All authors have given approval to the final version of the manuscript.

### Acknowledgments

The authors thank ALBA synchrotron (Proposal 20250340022) for providing beam time for the HP AD-XRD experiments. The research leading to this result has been cofunded by the project NEPHEWS under Grant Agreement No. 101131414 from the EU Framework Programme for Research and Innovation Horizon Europe. N.B. and D.E. thank the Generalitat Valenciana for the Postdoctoral Fellowship No. CIAPOS/ 2024/235 and D.E. and P.Z. thank the financial support of Generalitat Valenciana through grant CIAPOS/2024/383. G.B acknowledges the Generalitat Valenciana for the postdoctoral fellowship CIAPOS/2023/346. BJK Acknowledges the support of the Australian Research Council (DP230100558) for the work undertaken at the University of Sydney.